\documentclass[twocolumn,pra,amsmath,superscriptaddress,showpacs]{revtex4}
\usepackage{amstext}
\usepackage{amsmath}            
\usepackage{amssymb}            
\usepackage{graphicx}           
\usepackage{latexsym}
\usepackage{bm}

\begin{document}
\title  { Kraus representation of damped harmonic oscillator and its application}
\author{Yu-xi Liu}
\affiliation{The Graduate University for Advanced Studies
 (SOKENDAI), Shonan Village, Hayama, Kanagawa 240-0193, Japan}
\affiliation{Frontier Research System,  The Institute of Physical
and Chemical Research (RIKEN), Wako-shi 351-0198, Japan}
\author{\c{S}ahin K. \"Ozdemir}
 \affiliation{The Graduate University for Advanced Studies
 (SOKENDAI), Shonan Village, Hayama, Kanagawa 240-0193, Japan}
 \affiliation{SORST Research Team for Interacting Career Electronics,
  Hayama, Kanagawa 240-0193, Japan}
   \affiliation{CREST Research Team for Photonic Quantum Information, Hayama,
Kanagawa 240-0193, Japan}
\author{Adam Miranowicz}
 \affiliation{The Graduate University for Advanced Studies
 (SOKENDAI), Shonan Village, Hayama, Kanagawa 240-0193, Japan}
 \affiliation{SORST Research Team for Interacting Career Electronics,
  Hayama, Kanagawa 240-0193, Japan}
 \affiliation{Nonlinear Optics Division, Physics Institute, Adam
 Mickiewicz University, 61-614 Pozna\'n, Poland}

\author{Nobuyuki Imoto}
 \affiliation{The Graduate University for Advanced Studies
 (SOKENDAI), Shonan Village, Hayama, Kanagawa 240-0193, Japan}
 \affiliation{SORST Research Team for Interacting Career Electronics,
  Hayama, Kanagawa 240-0193, Japan}
 \affiliation{CREST Research Team for Photonic Quantum Information, Hayama,
Kanagawa 240-0193, Japan}
\affiliation{NTT Basic Research
Laboratories, NTT Corporation, 3-1 Morinosato Wakamiya, Atsugi,
Kanagawa 243-0198, Japan}

\begin{abstract}
By definition, the Kraus representation of a harmonic oscillator
suffering from the environment effect, modeled as the amplitude
damping or the phase damping, is directly given by a simple
operator algebra solution. As examples and applications, we first
give a Kraus representation of a single qubit whose computational
basis states are defined as bosonic vacuum and single particle
number states. We further discuss the environment effect on qubits
whose computational basis states are defined as the bosonic odd
and even coherent states. The environment effects on entangled
qubits defined by two different kinds of computational basis are
compared with the use of fidelity.

\end{abstract}

\maketitle \pagenumbering{arabic}
\section{introduction}
Since the fast quantum algorithms were developed, the quantum
information and computation have aroused great enthusiasm among
physicists because of their potential applications. The basic unit
of the classical information is a bit, which can only have a
choice between two possible classical states denoted by $0$ or
$1$. The information carriers  in the quantum communication and
computation are called quantum bits or qubits. The main difference
between the qubit and the classical bit is that the qubit is
defined as a quantum superposition of two orthogonal quantum
states usually written as $|0\rangle$ and $|1\rangle$ called as
the computational basis states. It means that the qubit has an
infinite number of choices of the  states
$\alpha|0\rangle+\beta|1\rangle$ with the condition
$|\alpha|^2+|\beta|^2=1$. It is obvious that the superposition
principle allows application of quantum states to simultaneously
represent many different numbers, this is called the quantum
parallelism, which  enables the quantum computation to solve some
problems, such as factorization, intractable on a classical
computer.

However, a pure quantum superposition state is very fragile. A
quantum system cannot be isolated from the environment, it is
always open and interacts with the uncontrollable environment.
This unwanted interaction induces  entanglement between the
quantum system and the environment such that the pure
superposition state is destroyed, which results in an inevitable
noise in the quantum computation and information processing.
Unlike a closed system, whose final state $\rho^{\prime}$  can be
obtained by a unitary transformation $U$ of the initial state
$\rho$  as $\rho^{\prime}=U\rho U^{\dagger}$, the final state of
an open system  cannot  be described by a unitary transformation
of the initial state. Quantum operation formalism is usually used
to describe the behavior of an open quantum system in which the
final state $\rho^{\prime}$ and the initial state $\rho$  can be
related by a quantum operation ${\cal E}$ as $\rho^{\prime}={\cal
E}(\rho)$. In general, the quantum operation ${\cal E}$ on the
state $\rho$ can be described by the Kraus operator-sum
formalism~\cite{Kraus,MI} as ${\cal E}(\rho)=\sum_{\mu}E_{\mu}\rho
E^{\dagger}_{\mu}$ where the operation elements
$E_{\mu}(E^{\dagger}_{\mu})$ satisfy the completeness relation
$\sum_{\mu}E^{\dagger}_{\mu}E_{\mu}= I$. Operators $E_{\mu}$ act
on the Hilbert space $H_{S}$ of the system. They can be expressed
as $E_{\mu}=\langle e_{\mu}|U_{t}|e_{0}\rangle$ by the total
unitary operator $U_{t}$ of the system and the environment with an
orthonormal basis $|e_{\mu}\rangle$ for the Hilbert space $H_{E}$
and the initial state $\rho_{E}=|e_{0}\rangle\langle e_{0}|$ of
the environment. This elegant representation is extensively
applied to describe the quantum information processing
~\cite{B,N}.

In practice, the environment effect on the quantum system is a
more complicated problem. There are two  ideal models of noise
called the amplitude and phase damping, which can capture many
important features of the noise~\cite{MI}. They are applied in
many concrete discussions to model noise of the quantum
information processing, e.g., reference~\cite{lidar}.  As a
system, a single-mode harmonic oscillator is the simplest and most
ideal model to represent a single mode light field, vibration
phonon mode, or excitonic wave. For convenience, we refer this
harmonic oscillator to a single-mode light field in this paper. A
study of a single harmonic oscillator suffering from damping is
expected to give us an easier grasp of the nature of damping. The
Kraus representation of a harmonic oscillator suffering from the
above two kinds of noise has been given by Chuang {\em et al.},
simply modeling environment as a single mode
oscillator~\cite{MI,isaac}. However generally speaking, the
environment is usually  described as a system of multimode
oscillators for both the phase damping and amplitude damping.
Milburn {\em et al.} have also given the Kraus representation of a
harmonic oscillator suffering from the amplitude damping by
modeling the environment as a system of the multimode oscillators
using the quantum measurement theory~\cite{milburn}. Although
there are many studies about Kraus representation, e.g. in
Refs.~\cite{Kraus,lidar,kraus2},  to the best of our knowledge,
there is no proof of the Kraus representation for a harmonic
oscillator suffering from amplitude damping or phase damping by
directly using its definition when the environment is
 modeled as a system of multimode oscillators. In view of the
importance of the Kraus representation in the quantum information,
we would like to revisit this question in this paper. We will
extend the proof of Chuang {\em et al.}, that is, we will give the
Kraus representation using a simple operator algebra method by
modeling the system as a single mode oscillator and the
environment as multimode oscillators according to the definition
of the Kraus representation.

Our paper is organized as follows. In Sec. II, we give the Kraus
representation for the amplitude damping case, then we discuss the
effect of the environment on a qubit whose computational basis is
defined by odd and even coherent states with decay of at most one
particle, we compare this result with that of a qubit defined by
vacuum and single photon number states. In Sec. III, we give the
Kraus representation for the phase damping case, and explain its
effect. Similar to the analysis in Sec. II, we also compare the
effect of the phase damping on the entangled qubit defined by
different states. Finally, we give our conclusions in Sec. IV.

\section{amplitude damping}\label{section3}
\subsection{Kraus representation}
One of the most important reasons for the quantum state change is
the energy dissipation of the system induced by the environment.
This energy dissipation can be characterized by an amplitude
damping model. Ideally we can model the system, for example, a
single mode cavity field, as a harmonic oscillator, which is
coupled to the environment  modeled as multimode oscillators. Then
the whole Hamiltonian of the system and the environment can be
written as
\begin{equation}
H=\hbar\omega b^{\dagger}b+\hbar\sum_{i}\omega_{i}
b_{i}^{\dagger}b_{i}+\hbar\sum_{i}
g_{i}(b^{\dagger}b_{i}+bb^{\dagger}_{i})\label{eq:1}
\end{equation}
under the rotating wave approximation, where $b (b^{\dagger})$ is
the annihilation (creation) operator of the harmonic oscillator of
frequency $\omega$, and $b_{i} (b^{\dagger}_{i})$ is the
annihilation (creation) operator of the $i$th mode of the
environment with frequency $\omega_{i}$, $g_{i}$ is coupling
constant between the system and  the $i$th mode of the
environment. We assume that the initial state of the whole system
is
\begin{equation}
\rho(0)=\rho_{S}(0)\otimes\rho_{E}(0),
\end{equation}
where $\rho_{S}(0)$ and $\rho_{E}(0)$ denote the initial states of
the system and environment respectively. With the time evolution,
the initial state evolves into
\begin{equation}
\rho(t)=U(t)\rho_{S}(0)\otimes\rho_{E}(0)U^{\dagger}(t)
\end{equation}
with $U=\exp(-iHt/\hbar)$. Since we are only interested in the
time evolution of the system, we perform a partial trace over the
environment and find the reduced density operator of the system as
\begin{eqnarray}
&&\rho_{S}(t)={\rm Tr}_{E}\left\{
U(t)\rho_{S}(0)\otimes\rho_{E}(0)U^{\dagger}(t)\right\} \label{4}\\
&&=\sum_{\{k_{i}\}}\langle
\{k_{i}\}|U(t)\rho_{S}(0)\otimes\rho_{E}(0)U^{\dagger}(t)|\{k_{i}\}\rangle
\nonumber
\end{eqnarray}
where $\{|\{k_{i}\}\rangle\equiv|k_{1}\cdots
k_{i}\cdots\rangle=\Pi_{i} \otimes|k_{i}\rangle \}$ is an
orthonormal basis of the Hilbert space $H_{E}$ for the
environment, and $k_{i}$ denotes that there are $k_{i}$ bosonic
particles in the $i$th mode. We assume that the environment is at
 zero temperature, so its initial state is in the vacuum state
$|\{0\} \rangle\equiv \Pi_{i} \otimes|0_{i}\rangle=|\cdots
0\cdots\rangle$ of multimode harmonic oscillators. The Hilbert
space $H_{E}$ can be decomposed into a direct sum of many
orthogonal subspaces $H_{k}$ as
$H_{E}=\bigoplus_{k=0}^{\infty}H_{k}$ with an orthonormal basis
{$\{|\{k_{i}\}\rangle \}$}, which satisfies the condition
$\sum_{i=1}^{\infty}k_{i}=k$  for each subspace $H_{k}$. So we can
regroup the summation in Eq.(\ref{4}) and define the quantum
operation element $A_{k}$ as follows
\begin{equation}
A_{k}(t)=\sum_{\{k_{i}\} }^{k}{'} \ \langle
\{k_{i}\}|U(t)|\{0\}\rangle
\end{equation}
which acts on the Hilbert space $H_{S}$ of the system. Hereafter,
$\sum{'}$ stands for summation under the condition
$\sum_{i}k_{i}=k$; $A_{k}(t)$ means that there are $k$ bosonic
particles  absorbed by the environment through the evolution over
a finite time $t$, and $A^{\dagger}_{k}(t)$ denotes the Hermitian
conjugate of $A_{k}(t)$. Then Eq.(\ref{4}) can be rewritten by
virtue of the operation elements $A_{k}(t)$ and its Hermitian
conjugate $A^{\dagger}_{k}(t)$ as a concise form called  the Kraus
representation~\cite{MI}
\begin{eqnarray}
\rho_{S}(t)=\sum_{k=0}^{\infty}A_{k}(t)\rho_{S}(0)
A^{\dagger}_{k}(t).\label{6}
\end{eqnarray}
It is very easy to check that the elements $A_{k}(t)$ satisfy the
relation $\sum_{k}A^{\dagger}_{k}(t)A_{k}(t)=I$. In order to
further give the solution of the operator $A_{k}(t)$, we first use
the number state $|n\rangle$ of the system to define an
orthonormal basis $\{|n\rangle\}$ of the Hilbert space $H_{S}$ of
the system. Then the matrix representation of the operator
$A_{k}(t)$ can be written as
\begin{eqnarray}
A_{k}(t)=\sum_{m,n} A_{m,n}^{k}(t)|m\rangle\langle n|\label{7}
\end{eqnarray}
 with the matrix  element
\begin{equation}
A_{m,n}^{k}(t)=\sum_{\{k_{i}\} }^{k}{'}\ \langle m|\langle
\{k_{i}\}|U(t)|\{0\}\rangle|n\rangle. \label{8}
\end{equation}
Because the state $U(t)|\{0\}\rangle|n\rangle$ can be written as
$U(t)(b^{\dagger})^n/\sqrt{n!}|\{0\}\rangle|0\rangle=
[b^{\dagger}(-t)]^{n}/\sqrt{n!}|\{0\}\rangle|0\rangle$. Using the
Wigner-Weisskopf approximation~\cite{ll}, the operator
$b^{\dagger}(t)$ can be obtained by solving the Heisenberg
operator equation of motion for Hamiltonian (\ref{eq:1}) as
\begin{equation}
b^{\dagger}(t)=u(t)b^{\dagger}(0)+
\sum_{i=1}^{\infty}v_{i}(t)b^{\dagger}_{i}(0),\label{eq:9}
\end{equation}
where $u(t)=\exp{(-\frac{\gamma}{2}t+i\omega t)}$ and the Lamb
shift has been neglected.  The damping rate $\gamma$ is defined as
$\gamma=2\pi\epsilon(\omega)|g(\omega)|^2$ with  a spectrum
density $\epsilon(\omega)$ of  the environment. The details of the
time dependent parameter $v_{i}(t)$ can be found in
~\cite{ll,sun}. We can use Eq.(\ref{eq:9}) to expand
$U(t)|\{0\}\rangle|n\rangle$ as many terms characterized by the
particle number $k$ lost from the system, and each term with fixed
$k$ can be expressed as
\begin{eqnarray}
\left[u^{*}(t)\right]^{n-k}&&\sqrt{\frac{1}{k!}\left(
\begin{array}{c}n \\k
\end{array}\right)}
\sum_{i_{1}\cdots\neq i_{q}\cdots \neq i_{k}=0}^{\infty}
\sum_{\{l_{i}\}}^{k}{'}\,
\frac{k!}{l_{1}!\cdots l_{k}!}\nonumber\\
&&\times \prod_{q=1}^{k}
\left(v^{*}_{i_{q}}(t)b^{\dagger}_{i_{q}}(0)\right)^{l_{q}}|\{
0\}\rangle|n-k\rangle. \label{9}
\end{eqnarray}
By using Eqs.(\ref{7})-(\ref{9}), we can obtain the product of the
matrix elements of $A^{k}_{m,n}$ and its Hermitian conjugate as
\begin{equation}
A^{k}_{m,n}({A^{k}}_{m,n})^{\dagger}=
\left(\begin{array}{c}n\\k\end{array}\right)
[\eta(t)]^{(n-k)}[1-\eta(t)]^{k}\delta_{m,n-k}
\end{equation}
with $\eta(t)=e^{-\gamma t}$,  where we have used the completeness
of the subspace with fixed $k$
\begin{equation}
\sum_{\{k_{i}\}}^{k}{'}\ | \{k_{i}\}\rangle\langle \{k_{i}\}|=1,
\end{equation}
and condition $|u(t)|^2+\sum_{i}|v_{i}(t)|^2=1$, which comes from
the commutation relation $\left[ b(t), b^{\dagger}(t) \right]=1$.
It is clear that  the non-zero matrix element $A^{k}_{m,n}(t)$
must satisfy the relation $n-m=k$, and all contributions to the
elements of the quantum operation $A_{k}(t)$ come from the states
$|n\rangle$ satisfying the condition $n\geq k$.  Then we can take
each matrix element $A^{k}_{m,n}(t)$ as a real number and finally
obtain~\cite{isaac,milburn}
\begin{equation}\label{133}
 A_{k}(t)=\sum_{n=k}^{\infty}
\sqrt{\left(\begin{array}{c}n\\k\end{array}\right)}
[\eta(t)]^{(n-k)/2}[1-\eta(t)]^{k/2}|n-k\rangle\langle n|.
\end{equation}
It is not difficult to prove  that
$\sum_{k}A^{\dagger}_{k}(t)A_{k}(t)=1$. In Eq.(\ref{133}), one
finds that$(1-\eta(t))^{k/2}$ is the probability that the system
loses $k$ particles up to time $t$, or the probability that the
state $|n\rangle$ is undecayed corresponds to
$[\eta(t)]^{(n-k)/2}$ for $k$ particle decay process up to time
$t$. For the convenience, we will write $\eta(t)$ as $\eta$ in the
following expressions.

Using a single qubit $\rho$ defined by the bosonic number state
$\{|0\rangle, \hspace{0.1mm} |1\rangle\}$ as the computational
basis states, we can simply demonstrate how to give its Kraus
representation when it suffers from the amplitude damping. By
Eq.(\ref{133}), it is very easy to find that the quantum operation
on qubit $\rho$ includes only two operational elements $A_{0}$ and
$A_{1}$, that is

\begin{equation}\label{eq:14}
{\cal E}(\rho)=A_{0}\rho A^{\dagger}_{0}+A_{1}\rho A^{\dagger}_{1}
\end{equation}
with
\begin{subequations}\label{eq:166}
\begin{eqnarray}
A_{0}(t)&=& |0\rangle\langle 0|+\sqrt{\eta}|1\rangle\langle 1|,\\
A_{1}(t)&=& \sqrt{1-\eta}|0\rangle\langle 1|,
\end{eqnarray}
\end{subequations}
where $\sqrt{1-\eta}$ is the probability that the system lose one
particle up to time $t$.

\subsection{Effect of amplitude damping on qubits}
In this subsection, we will demonstrate  an application of the
above conclusions about the Kraus representation. We know that two
kinds of logical qubits, whose computational basis states are
defined by using the bosonic even and odd coherent
states~\cite{haroche} or vacuum and bosonic single particle number
states, are accessible in experiments. For brevity, we refer to
qubits defined via the even and odd coherent states as the
(Schr\"odinger) cat-state qubits by contrast to the Fock-state
qubits defined by vacuum and single-photon states in the following
expressions. It has been shown that the bit flip errors caused by
a single decay event, which results from the spontaneous
emissions, are more easily corrected by a standard error
correction circuit for the logical qubit defined by the bosonic
even and odd coherent states~\cite{ptc} than that defined by the
bosonic vacuum and single particle number states. We know that the
entangled qubit takes an important role in the quantum information
processing. So in the following, we will study the effect of the
environment on entangled qubits defined by the above two kinds of
computational basis states when they are subject to at most one
decay event caused by the amplitude damping. Let $|\alpha\rangle$
be a bosonic coherent state (we denote this single bosonic mode as
a single-mode light field, but it can be generalized to any
bosonic mode, for example, excitonic mode, vibrational mode of
trap ions and so on). In general, if a single-mode light field,
which is initially in the coherent state $|\alpha\rangle$, suffers
from amplitude damping, the Kraus operation element (\ref{133})
changes it as follows
\begin{eqnarray}
A_{k}|\alpha\rangle=e^{-\frac{(1-\eta)|\alpha|^2}{2}}
\frac{(\alpha\sqrt{1-\eta})^{k}}{\sqrt{k!}}
|\sqrt{\eta}\alpha\rangle,
\end{eqnarray}
and the normalized state, which is denoted by ${\cal
N}[A_{k}|\alpha\rangle]$, can be written as
\begin{eqnarray}
{\cal N}[A_{k}|\alpha\rangle]=|\sqrt{\eta}\alpha\rangle,
\end{eqnarray}
which means that the coherent state remains coherent under the
amplitude damping, but its amplitude $\alpha$ is reduced to
$\sqrt{\eta}\alpha$ due to the interaction with the environment.

Now we define a logical zero-qubit state $|0\rangle_{L}$ and a
logical one-qubit state $|1\rangle_{L}$ using the even and odd
coherent states as
\begin{eqnarray}
|0\rangle_{L}&=&N_{+}(|\alpha\rangle+|-\alpha\rangle)\nonumber\\
|1\rangle_{L}&=&N_{-}(|\alpha\rangle-|-\alpha\rangle)\label{16}
\end{eqnarray}
with $N_{\pm}=(2\pm 2e^{-2|\alpha|^2})^{-\frac{1}{2}}$ and
subscript $L$ denotes the logical state. If we know that the
system loses at most one photon with the time evolution but we do
not have the detailed information on the lost photons. Then, we
only need to calculate two Kraus operators $A_{0}$ and $A_{1}$
corresponding to no-photon and single-photon decay events,
respectively.

No-photon decay event changes the  logical qubits $|0\rangle_{L}$
and $|1\rangle_{L}$ as follows
\begin{mathletters}
\begin{eqnarray}
|\widetilde{0}\rangle&=&{\cal N}[A_{0}|0\rangle_{L}]=
N^{\prime}_{+}\left(|\sqrt{\eta}\alpha\rangle+
|-\sqrt{\eta}\alpha\rangle \right),\\
|\widetilde{1}\rangle&=&{\cal
N}[A_{0}|1\rangle_{L}]=N^{\prime}_{-}\left(|\sqrt{\eta}\alpha\rangle-
|-\sqrt{\eta}\alpha\rangle \right),
\end{eqnarray}
\end{mathletters}
where $|\widetilde{0}\rangle$  and $|\widetilde{1}\rangle$ are
normalized states of $A_{0}|0\rangle_{L}$ and $A_{0}|1\rangle_{L}$
with the normalized constants
$N^{\prime}_{\pm}=\left\{2\pm2\exp(-2\eta|\alpha|^2)\right\}^{-1/2}$.
It is obvious that no-photon decay event  reduces only the
intensity of the logical qubit, leaving the even and odd
properties of the logical qubit states unchanged. A simple
analysis of how the noisy channel affects the original quantum
state can be made by calculating the fidelity defined as the
average value of the final reduced density matrix $\rho(t)$ with
initial pure state $|\psi\rangle$, that is
$f(t)=\langle\psi|\rho|\psi\rangle$. The fidelity $f_{i}(t)$ of
the qubit state (\ref{16}) with no-photon decay event can be
obtained as
\begin{equation}
f_{i}(t)=\frac{(e^{\sqrt{\eta}|\alpha|^2}+(-1)^ie^{-\sqrt{\eta}|\alpha|^2})^2}
{(e^{|\alpha|^2}+(-1)^ie^{-|\alpha|^2})(e^{\eta|\alpha|^2}+(-1)^ie^{-\eta|\alpha|^2})}
\end{equation}
with $i=0, \ 1 $, which denotes zero-qubit state or one-qubit
state. But for a single-photon decay event, we can have the
following
\begin{eqnarray}
|\widetilde{1}\rangle={\cal N}[A_{1}|0\rangle_{L}],\hspace{1cm}
|\widetilde{0}\rangle={\cal N}[A_{1}|1\rangle_{L}].
\end{eqnarray}
We find that the single photon decay event flips an even coherent
state to an odd coherent state with the reduction of the amplitude
and vice versa. It is very easy to prove that
\begin{mathletters}
\begin{eqnarray}
|\widetilde{1}\rangle&=&{\cal N}[A_{(2m+1)}|0\rangle_{L}]={\cal
N}[A_{2m}|1\rangle_{L}],\\
|\widetilde{0}\rangle&=&{\cal N}[A_{(2m+1)}|1\rangle_{L}]={\cal
N}[A_{2m}|0\rangle_{L}]
\end{eqnarray}
\end{mathletters}
with $m=0,\ 1,\ 2 \cdots $,  which means that the even number of
photons decay event only reduces the intensity of the logical
signal, and keeps the even and odd properties of the logical state
unchanged; but the odd number of photons decay event flips a qubit
causing error and reduces the intensity of the qubit signal. If
the coherent amplitude of the cat states is infinitely large, that
is, $|\alpha|\rightarrow\infty$, then we find that the fidelity
$f_{i}(t)\approx 1$ with a larger $\eta$, so under this condition,
we can say that the no-photon decay event essentially leaves the
logical qubit states unchanged and the single-photon decay event
causes a qubit flip error. It means that the use of the even and
odd coherent states as logical qubit states $|0\rangle_{L}$ and
$|1\rangle_{L}$ is better than the use of the vacuum state
$|0\rangle$ and single-photon state $|1\rangle$ as logical qubit
states in the amplitude damping channel with a few photons loss.
Because the single-photon decay event changes vacuum  and
single-photon states to $A_{1}|0\rangle=0$ and
$A_{1}|1\rangle=|0\rangle$, the single-photon decay event is an
irreversible process for vacuum and single-photon states.

An entangled pair of qubits, whose computational basis states are
defined by the even and odd coherent states (\ref{16}), can be
written as
\begin{equation}\label{eq:266}
|\Psi\rangle=\frac{1}{\sqrt{2}}\left\{|0\rangle_{L}|1\rangle_{L}+|1\rangle_{L}|0\rangle_{L}\right\}.
\end{equation}
On the basis of the above discussions, we can study the effect  of
environments on entangled qubits when each mode loses at most one
photon. For simplicity, two modes are assumed to suffer from
effect of two same independent environments. There is no direct
interaction between two systems. After the environment performs
the following four measurements $\{E_{0}=A_{0}\otimes
A_{0},\hspace{0.1cm} E_{1}=A_{0}\otimes A_{1},\hspace{0.1cm}
E_{2}=A_{1}\otimes A_{0}, \hspace{0.1cm} E_{3}=A_{1}\otimes
A_{1}\}$ with $A_{0}$ an $A_{1}$ determined by Eq.(\ref{133}), the
entangled qubit (\ref{eq:266}) becomes the following mixed state
\begin{eqnarray}\label{eq:16}
\rho (t) &=&\frac{\sum_{i=0}^{3}E_{i}\rho (0)E_{i}^{\dagger
}}{{\rm Tr} \left\{ \sum_{i=0}^{3}E_{i}\rho (0)E_{i}^{\dagger
}\right\} }=\tilde{N} \left\{ 2p|\alpha |^{2}\left(
\frac{N_{-}^{\prime 2}}{N_{+}^{\prime 2}}|\tilde{0}\tilde{0}
\rangle \langle \tilde{0}\tilde{0}| \right. \right.  \nonumber \\
&& \left. +\frac{N_{+}^{\prime 2}}{N_{-}^{\prime 2}}
|\tilde{1}\tilde{1}\rangle \langle
\tilde{1}\tilde{1}|+|\tilde{0}\tilde{0} \rangle \langle
\tilde{1}\tilde{1}|+|\tilde{1}\tilde{1}\rangle \langle
\tilde{0}\tilde{0}|\right) \nonumber \\
&&\left. +(1+p^{2}|\alpha |^{4})(|\tilde{0}\tilde{1}\rangle
+|\tilde{1}\tilde{0} \rangle )(\langle \tilde{0}\tilde{1}|+\langle
\tilde{1}\tilde{0}|) \right\}
\end{eqnarray}
with $p=1-\eta$ and the normalized constant
\begin{eqnarray}
\widetilde{N}^{-1}&=&2\frac{(1+p|\alpha|^2)^2-e^{-4\eta|\alpha|^2}(1-p|\alpha|^2)^2}
{1-e^{-4\eta|\alpha|^2}}.
\end{eqnarray}
 The fidelity $F_{1}(t)$ of the entangled qubit $|\Psi\rangle$
with at most one photon decay for each mode can be calculated as
\begin{eqnarray}\label{eq:29}
F_1(t)&=&\langle\Psi|\rho(t)|\Psi\rangle \label{23}\\
&=&\frac{\left( 1 + |\alpha|^4p^2 \right) {\rm
csch}(2|\alpha|^2){{\rm sh}^2 (2\sqrt{\eta}|\alpha|^2)}}
{2|\alpha|^2p \ {\rm ch} (2|\alpha|^2\eta) + \left( 1 +
|\alpha|^4p^2 \right) {\rm sh} {(2|\alpha|^2\eta)}} \nonumber,
\end{eqnarray}
If each mode can dissipate any number of photons, but we know
nothing about the details of dissipation, then we must sum up all
possible environment measurements on the system, obtaining the
reduced matrix operator of the system as
\begin{eqnarray}
&&\rho^{\prime}(t)=\frac{1}{2-2e^{-4|\alpha|^2}}\left\{|\beta\rangle|\beta\rangle\langle\beta|\langle\beta|
+|-\beta\rangle|-\beta\rangle\langle-\beta|\langle-\beta|\right\}\nonumber\\
&&-\frac{M}{2-2e^{-4|\alpha|^2}}
\left\{|-\beta\rangle|-\beta\rangle\langle\beta|\langle\beta|
+|\beta\rangle|\beta\rangle\langle-\beta|\langle-\beta|\right\}
\end{eqnarray}
with $M=e^{-4|\alpha|^2(1-e^{-\gamma t})}$ and
$|\beta\rangle=|u(t)\alpha\rangle$. Then we obtain the fidelity
$F_{2}(t)$ as
\begin{equation}
F_2(t)={\rm ch}(2|\alpha|^2p)\,{{\rm csch}^2(2|\alpha|^2)}\, {{\rm
sh}^2(2\sqrt{\eta}|\alpha|^2)}.\label{25}
\end{equation}
We can find that the logical state $|0\rangle_{L}$ and
$|1\rangle_{L}$ in Eq.(\ref{eq:266}) can be reduced to the vacuum
state $|0\rangle$ and single-photon state $|1\rangle$,
respectively in the limit of the weak light field
$|\alpha|\rightarrow 0$. It is easy to check that
\begin{mathletters}
\begin{eqnarray}
A_{0}|0\rangle&=&|0\rangle,\hspace{0.5cm}
A_{0}|1\rangle=\sqrt{\eta}|1\rangle,\\
A_{1}|0\rangle&=&0,\hspace{0.8cm}
A_{1}|1\rangle=\sqrt{1-\eta}|0\rangle
\end{eqnarray}
\end{mathletters}
which means that the $|0\rangle$ qubit state is invariant when
no-photon decay event happens, but the amplitude of the
$|1\rangle$ qubit state is reduced to $\sqrt{\eta}$. If the system
leaks one photon, then the $|0\rangle$ qubit state vanishes and
the $|1\rangle$ qubit state returns to the $|0\rangle$ with the
probability $\sqrt{1-\eta}$. The entangled qubit, whose
computational basis states are defined by vacuum and single-photon
states, can be written as
\begin{equation}
|\Psi\rangle=\frac{1}{\sqrt{2}}(|0\rangle|1\rangle+|1\rangle|0\rangle)\label{19}
\end{equation}
which can be obtained by setting $|\alpha|\rightarrow 0$ in
Eq.(\ref{eq:266}). We can use the same step to obtain the fidelity
$F^{\prime}_{1}(t)$  corresponding to Eq.(\ref{19}) when both
modes are subject to the amplitude damping as
\begin{equation}\label{eq:35}
F^{\prime}_{1}(t)=\eta(t),
\end{equation}
where we assume that the damping is the same for the two modes.
When the weak light field limit $|\alpha|\rightarrow 0$ is taken,
we find that  Eq.(\ref{23}) and Eq.(\ref{25}) can be written as
\begin{eqnarray}
F_{1}(t)&=&\eta+\frac{\eta}{3}(1-5\eta+3\eta^2+\eta^3)
|\alpha|^4+{\cal O}(|\alpha|^8), \nonumber\\
F_{2}(t)&=&\eta+\frac{2}{3}\eta(1-4\eta+3\eta^2)|\alpha|^4+{\cal
O}(|\alpha|^8).
\end{eqnarray}
It is clear that all the higher order terms of $|\alpha|^2$ can be
neglected in the limit $|\alpha|\rightarrow 0$, so Eq.(\ref{23})
and Eq.(\ref{25}) can return to $F^{\prime}_{1}(t)$ in the weak
field approximation.
\begin{figure}
\includegraphics[width=9cm]{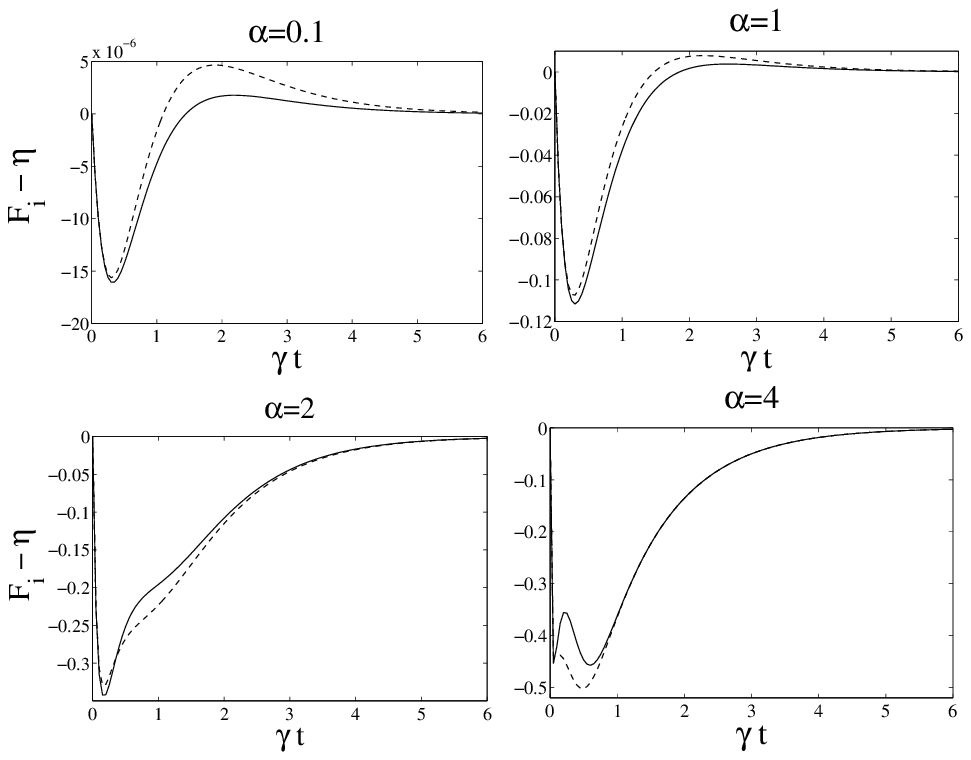}
\caption[]{Difference of fidelities: $F_1-\eta$ (solid curves),
given by Eqs. (\ref{eq:29}) and (\ref{eq:35}), and  $F_2-\eta$
(dashed curves), given by Eqs. (\ref{25}) and (\ref{eq:35}),
versus scaled time $\gamma t$ for different values of amplitude
$\alpha$. }\label{fig1}
\end{figure}
By recalculating the difference
\begin{equation}
F_1(t)-F_2(t)=-\frac{1}{3}\eta (1-\eta)^3 |\alpha|^4+{\cal
O}(|\alpha|^8),
\end{equation}
it is clear that fidelity $F_1(t)$ is less than $F_2(t)$ in the
low-intensity regime. These predictions are confirmed by the exact
evolution of fidelities depicted in Fig. (\ref{fig1}) for
$|\alpha|\le 1$. To compare the fidelities in the short-time
evolution, we expand the difference of Eqs. (\ref{23}) and
(\ref{25}) in power series in $\Gamma t$ as follows
\begin{eqnarray}
F_1(t)-F_2(t)&=&-\frac{2}{3}|\alpha|^6 {\rm coth}(2|\alpha|^2)
(\gamma t)^3 +{\cal O}\{(\gamma t)^4\} \nonumber \\
\end{eqnarray}
which shows that $F_1(t)$ is less than $F_2(t)$ for arbitrary
intensities in the short-time regime although the difference might
be very small as presented in Fig. (\ref{fig1}) for
$|\alpha|=2,4$. In the long-time limit, we can expand the
difference in a series of $\eta$ resulting in
\begin{eqnarray}
F_1(t)-F_2(t)&=&\frac{2|\alpha|^2}{{\rm sh}(2|\alpha|^2)}
[1+|\alpha|^4-2|\alpha|^2{\rm coth}(2|\alpha|^2)]\eta \nonumber
\\&&+{\cal O}\{\eta^2\}
\end{eqnarray}
which is negative only for the intensity
$|\alpha|^2<1.1997\cdots$.  For higher intensity, we find that
$|\widetilde{1}\rangle\approx|1\rangle_{L}$ and
$|\widetilde{0}\rangle\approx|0\rangle_{L}$, and $p=1-\eta$ will
become small with the time evolution, then the fidelity $F_{1}(t)$
will become greater than the fidelity $F_{2}(t)$ in some regimes.
But in the low intensity limit we find that
$|\widetilde{1}\rangle\approx|0\rangle$ and
$|\widetilde{0}\rangle\approx|0\rangle$,  then the fidelity
$F_{1}(t)$ will become less than $F_{2}(t)$.

\section{phase damping}
\subsection{Kraus representation}
A state can be a superposition of different states, which is one
of the main characteristics of the quantum mechanics. The relative
phase and amplitude of the superposed state determines the
properties of the whole state. If the relative phases of the
superposed states randomly change with the time evolution, then
the coherence of the quantum state will be destroyed even if the
eigenvalue of the quantum system will be changed. This kind of
quantum noise process is called the phase damping.  We can have
the Hamiltonian of a harmonic oscillator suffering from the phase
damping as
\begin{equation}
H=\hbar\omega b^{\dagger}b+\hbar\sum_{i}\omega_{i}
b_{i}^{\dagger}b_{i}+\hbar\sum_{i}
\chi_{i}b^{\dagger}b(b_{i}+b^{\dagger}_{i}),\label{eq:24}
\end{equation}
where $b (b^{\dagger})$ is the annihilation (creation) operator of
the harmonic oscillator with frequency $\omega$; and $b_{i}
(b^{\dagger}_{i})$ is the annihilation (creation) operator of the
$i$th mode of the environment with frequency $\omega_{i}$.
$\chi_{i}$ is a coupling constant between the system and the $i$th
mode of the environment. We can solve the Heisenberg operator
equation of motion, and very easily obtain the solution of the
system operator $b(t)$ as
\begin{mathletters}
\begin{eqnarray}
&&b(t)=b(0)\exp\left[-i\omega
t-i\sum\chi_{j}t(b^{\dagger}_{j}(t)+b_{j}(t))\right],\\
&&b_{j}(t)=b_{j}(0)\exp\left[-i\omega_{j}
t-i\chi_{j}tb^{\dagger}b\right].
\end{eqnarray}
\end{mathletters}
We  also assume that the system-environment is initially in a
product state $\rho(0)=\rho_{S}(0)\otimes\rho_{E}(0)$. We apply
the time-dependent unitary operator $U(t)=e^{-iHt/\hbar}$ with the
Hamiltonian $H$ determined by Eq.(\ref{eq:24}) to the state
$|\{0\}\rangle |m\rangle$ as follows
\begin{eqnarray}
&&U(t)|\{0\}\rangle |m\rangle=\frac{[b^{\dagger}(-t)]^ m
}{\sqrt{m!}}|\{0\}\rangle |0\rangle =\nonumber\\
&&e^{\left\{-im\omega t-i\sum\chi_{j}m
t(b^{\dagger}_{j}(-t)+b_{j}(-t))\right\}}\frac{b^{\dagger
m}(0)}{\sqrt{m!}}|\{0\}\rangle |0\rangle.\label{eq:26}
\end{eqnarray}
By assuming that the environment scatters off the quantum system
randomly into the states $\{|k_{1}\cdots k_{i}\cdots\rangle \}$
with the total particle number $k$, then the Kraus operators can
be defined as follows
\begin{equation}
P_{k}(t)=\sum_{\{k_{i}\} }^{k}{'}\ \langle
\{k_{i}\}|U(t)|\{0\}\rangle\label{27}
\end{equation}
with its Hermitian conjugate $P^{\dagger}_{k}$.Using Eq
(\ref{eq:26}) and Eq (\ref{27}),  the reduced density operator of
the system can be expressed as
\begin{equation}
\rho(t)=\sum_{k=0}^{\infty}P_{k}(t)\rho(0)P^{\dagger}_{k}(t)
\end{equation}
with
\begin{equation}
P_{k}(t)\equiv
P_{k}(\tau)=\sum_{n=0}^{\infty}\exp\left\{-\frac{1}{2}n^2\tau^2\right\}
\sqrt{\frac{\left(n^2\tau^2\right)^{k}}{k!}} |n\rangle\langle
n|\label{eq:27}
\end{equation}
where $\tau=t\sqrt{\Gamma}$ is a rescaled interaction time and
$\Gamma=\sum_{j}|\chi_{j}|^2$.
$\sqrt{1-\exp\left\{-n^2\tau^2\right\}}$ can be interpreted as the
probability that $n$ particles from the system are scattered by
the environment. It is clear that a sum of all $P_{k}$ satisfies
the condition $\sum_{k=0}^{\infty} P^{\dagger}_{k}(t)P_{k}(t)=I$.
We still use qubit as a simple example to investigate the phase
damping effect on it and give its Kraus representation. We assume
that our harmonic system is initially in
$|\psi\rangle=\alpha|0\rangle+\beta|1\rangle$ and suffers from the
phase damping. By virtue of Eq.(\ref{eq:27}), it is easy to find
that the quantum operation on the qubit $|\psi\rangle$ can be
expressed as
\begin{equation}
{\cal E}(\rho)=\sum_{k=0}P_{k}(\tau)\rho P^{\dagger}_{k}(\tau)
\end{equation}
with $\rho=|\psi\rangle\langle \psi|$,  and

\begin{eqnarray}
P_{k}(\tau)=\delta_{k0}|0\rangle\langle
0|+\exp\left\{-\frac{1}{2}\tau^2\right\}\frac{\tau^{k}}{\sqrt{k!}}
|1\rangle\langle 1|\label{29b},
\end{eqnarray}
where $\delta_{k0}$ is Kronecker delta. By contrast to the case of
the amplitude damping in which $k$ can only take $0$ and $1$,  in
the case of the phase damping, $k$ can take values from $0$ to
$\infty$, which means that the system can be scattered by  any
number of particles in the environment. In this sense, we can say
that there is an infinite number of quantum operation elements
$P_{k}(t)$ acting on the qubit $|\psi\rangle$ when it suffers from
a phase damping. We find that $P_{k\neq 0}(t)$  makes the state
$|0\rangle$ equal to zero, so we can rearrange  all operation
elements $P_{k\neq 0}(t)$ as one group, and redefine two Kraus
operators as
\begin{eqnarray}
E_{0}&=&|0\rangle\langle 0|+\exp\left\{-\frac{1}{2}\tau^{2}\right\}|1\rangle\langle 1|\nonumber\\
E_{1}&=&\sqrt{1-\exp\left\{-\tau^2\right\}}|1\rangle\langle 1|
\end{eqnarray}
which is the form of the reference~\cite{MI}. It is evident that
$E_{0}E^{\dagger}_{0}+E_{1}E^{\dagger}_{1}= I$, and $E_{0}$ means
that the environment returns to the ground state after it is
scattered by the system. Unlike the amplitude damping, $E_{1}$
means that the environment is scattered to another state, which is
orthogonal to its ground state. It does not mean that only one
particle in the environment is scattered by the system. We find
that the probability of  a photon from the system being scattered
by the environment is $\sqrt{1-\exp\left\{-\tau^2\right\}}$.

\subsection{Phase damping effect on qubits}
In this subsection, we will further discuss the phase damping
effect on a qubit whose computational basis states are defined by
the bosonic even and odd coherent states or vacuum and  single
photon number states. If the coherent state $|\alpha\rangle$ of
the system is  scattered by $k$ photons of the environment, then
it changes as
\begin{equation}
P_{k}(\tau)|\alpha\rangle=\frac{\tau^k}{\sqrt{k!}}
\sum_{n=0}^{\infty}\exp\{-\frac{1}{2}(n^2\tau^2+|\alpha|^2)\}\frac{n^k\alpha^n}{\sqrt{n!}}|n\rangle.
\end{equation}
We are interested in the change of the off-diagonal terms for the
system  state after it scatters the photons of the environment,
for $k$ photons scattering, we have

\begin{eqnarray}
&&P_{k}(\tau)|\alpha\rangle\langle\beta|P^{\dagger}_{k}(\tau)=\frac{\tau^{2k}}{k!}
\exp\{-\frac{1}{2}(|\alpha|^2+|\beta|^2)\}\\
&&\times\sum_{n,m=0}^{\infty}\exp\left\{-\frac{\tau^2}{2}(n^2+m^2)\right\}\frac{(nm)^k\alpha^n\beta^{*m}}{\sqrt{n!m!}}
|n\rangle\langle m| \nonumber.
\end{eqnarray}
 As the system can scatter an arbitrary number
of photons in the environment,  we need to sum all $P_{k}(\tau)$,
getting
\begin{eqnarray}
&&\sum_{k=0}^{\infty}P_{k}(\tau)|\alpha\rangle\langle\beta|P^{\dagger}_{k}(\tau)=
\exp\{-\frac{1}{2}(|\alpha|^2+|\beta|^2)\} \nonumber\\
&&\times\sum_{n,m}\exp\{-\frac{
\tau^2}{2}(n-m)^2\}\frac{\alpha^n\beta^{*m}}{\sqrt{n!m!}}
|n\rangle\langle m|.
\end{eqnarray}
In order to get a clear illustration of way how the phase damping
affects the logical qubit states, we can write out reduced density
matrix of logical states $|0\rangle_{L}$ and $|1\rangle_{L}$ in
Eq.(\ref{16}) as follows
\begin{mathletters}
\begin{eqnarray}
&&\rho_{0}(\tau)=\sum_{n,m}\frac{e^{-2
\tau^2(n-m)^2}\alpha^{2n}\alpha^{*2m}}{{\rm ch}|\alpha|^2\sqrt{(2n)!(2m)!}}|2n\rangle\langle 2m|\\
&&\rho_{1}(\tau)=\sum_{n,m}\frac{e^{-2
\tau^2(n-m)^2}\alpha^{2n+1}\alpha^{*2m+1}}{{\rm sh}
|\alpha|^2\sqrt{(2n+1)!(2m+1)!}}|2n+1\rangle\langle
2m+1|\nonumber\\
\end{eqnarray}
\end{mathletters}
It is obvious that when the qubit states are subject to phase
damping, although the even and odd properties are not changed, all
of the off-diagonal elements of the reduced matrix $\rho_{k}(t)$
with $(k=0, \ 1)$ tend to zero with the time evolution. In
contrast to  the amplitude damping, the qubit states suffering
from the phase damping are no longer pure states even though no
photon is lost. Because the relative phases of different
superposed elements of the coherent state have been destroyed by
the random scattering process of the environment. The fidelity
$f_{k}(t)$ ($k=0, \ 1$) of the logical qubit states (\ref{16}) can
be obtained as
\begin{equation}\label{eq:55}
f_{k}(\tau)=\frac{1}{{\rm ch}^{2}|\alpha
|^{2}-k}\sum_{n,m}\frac{|\alpha
|^{4(n+m+k)}e^{-2(n-m)^{2}\tau^{2}}}{(2n+k)!(2m+k)!},
\end{equation}
where $f_{k}(\tau)\leq 1 (k=0,1)$ with the equality sign holding
only in the limit $|\alpha|\rightarrow 0$ so that
$|0\rangle_{L}\approx |0\rangle$ and $|1\rangle_{L}\approx
|1\rangle$. But if we directly choose the vacuum state $|0\rangle$
and single photon state $|1\rangle$ as the logical qubit states,
and they independently go into the phase damping channel, the
fidelity is kept as one. It is because  the phase damping only
changes the off-diagonal terms. It means that in the phase damping
channel, the use of the vacuum state $|0\rangle$ and single-photon
state $|1\rangle$ as logical qubit states is better than the use
of the even and odd coherent states as logical qubit states
$|0\rangle_{L}$ and $|1\rangle_{L}$. The fidelities of the even
and odd coherent states suffering from phase damping are compared
in Fig. (\ref{fig2}). We find that in the low coherent intensity
of $|\alpha|^2$, the even coherent state can keep a better
fidelity than that of the odd coherent state. But the  fidelities
for the even and odd coherent states approach to each other with
increasing intensity, so the states have no difference for the
loss of information in the high coherent intensity.

To show this property analytically, we find that one summation in
Eq. (\ref{eq:55}) can be performed leading us to relation
($k=0,1)$
\begin{eqnarray}
f_{k} &=&\frac{1}{2({\rm ch}^{2}|\alpha |^{2}-k)}\left\{
I_{0}(2|\alpha
|^{2})+(-1)^{k}J_{0}(2|\alpha |^{2})\right.  \nonumber \\
&&+ 2\sum_{d=1}^{\infty }  e^{-2d^{2}\tau^{2}}\left.
[I_{2d}(2|\alpha |^{2})+(-1)^{k}J_{2d}(2x)]\right\}
\end{eqnarray}
where $J_{2d}(x)$ is the Bessel function and $I_{2d}(x)$ is the
hyperbolic Bessel function. If $|\alpha |^{2}\gg 1$ (on the scale
of Fig. 2, for $|\alpha |^{2}\geq 4$) then $I_{d}(2x)\gg
J_{d}(2x)$ and $($ch$^{2}|\alpha |^{2}-k)^{-1}\approx e^{-2|\alpha
|^{2}}$, which implies that both fidelities $f_{k}$ are
approximately equal to
\begin{equation}
f_{0}\approx f_{1}\approx 2e^{-2|\alpha |^{2}}\left(
I_{0}(2|\alpha |^{2})+2\sum_{d=1}^{\infty
}e^{-2d^{2}\tau^{2}}I_{2d}(2|\alpha |^{2})\right) .
\end{equation}
Assuming also long scaled interaction times,  $\tau^2\equiv\Gamma
t^{2}\gg 1$, the fidelities can be approximated by compact
formulas
\begin{eqnarray}
f_{0} \approx f_{1}&\approx& 2e^{-2|\alpha |^{2}}[I_{0}(2|\alpha
|^{2})+2e^{-2\tau^{2}}I_{1}(2|\alpha |^{2})]
\nonumber  \\
&\approx &\frac{1}{\sqrt{\pi }|\alpha |}\left[ 1+2\exp
(-2\tau^{2}) \right] .
\end{eqnarray}
\begin{figure}\label{fig2}
\includegraphics[width=6cm]{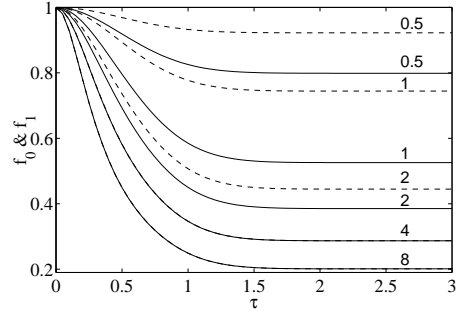}
\caption[]{Evolution of fidelities $f_0(\tau)$ (solid curves) and
$f_1(\tau)$ (dashed curves) are given by Eq. (\ref{eq:55})
 for different values of intensity $|\alpha|^2$. It
is seen that the curves for $f_0(\tau)$ and $f_1(\tau)$ coincide
if $|\alpha|^2\ge 4$. \label{fig2}}
\end{figure}

We can also study the effect of the phase damping on the entangled
qubit (\ref{eq:266}) based on the above discussions, and the
relevant calculations are straightforward. Here, we only give a
fidelity $F(t)$ for two modes subjected to the phase damping as
\begin{eqnarray} \label{eq:599}
F(\tau)&=&\frac{1}{2}f_{0}(\tau)f_{1}(\tau)+\frac{2e^{-\tau^2}}{{\rm
sh}^2(2|\alpha |^{2}) } \\ &&\times \left( \sum_{n,m}\frac{
|\alpha
|^{2(2n+2m+1)}e^{-2(n-m)(n-m+1)\tau^{2}}}{(2m)!(2n+1)!}\right)
^{2} \nonumber
\end{eqnarray}
where $f_{0}(\tau)$ and $f_{1}(\tau)$ are given by Eq.
(\ref{eq:55}). Assuming that both channels have the same phase
damping constant $\Gamma$. Eq. (\ref{eq:599}) simplifies to the
following single-sum formula
\begin{eqnarray}
F(\tau)
&=&\frac{1}{2}f_{0}(\tau)f_{1}(\tau)+\frac{2e^{-\tau^{2}}}{{\rm
sh}^{2}{
(2|\alpha |^{2})}} \\
&&\times \left( \sum_{d=1}^{\infty
}e^{-2d(d+1)\tau^{2}}I_{2d+1}(2{ |\alpha |^{2}})+I_{1}(2{|\alpha
|^{2}})\right) ^{2} \nonumber
\end{eqnarray}
which helps us to find approximate compact-form solutions.
E.g.,~for either long scaled interaction times, $\tau\gg 1$, or
low intensities, $|\alpha |^{2}\ll 1$, the fidelity $F(\tau)$
reduces to
\begin{eqnarray}
F(\tau)\approx
\frac{1}{2}f_{0}(\tau)f_{1}(\tau)+\frac{2e^{-\tau^{2}}}{{\rm sh}
^{2}{(2|\alpha |^{2})}}[I_{1}(2{|\alpha |^{2}})]^{2}.
\end{eqnarray}
On the other hand, by assuming high intensity, $\ {|\alpha
|^{2}}\gg 1$, together with $ \tau\gg 1$, one obtains the
following approximation
\begin{eqnarray}
F(\tau)\approx \frac{1}{2\pi {|\alpha |^{2}}}\left[ 1+4\exp
(-\tau^{2}) \right].
\end{eqnarray}

If  the qubit states are defined by the vacuum state $|0\rangle$
as logical zero state and single-photon state $|1\rangle$ as
logical one state, then for the entangled qubits (\ref{19}), if
both of them go into phase damping channels with the same damping
constant $\Gamma$, we can obtain the fidelity as
\begin{equation}\label{eq:58}
F^{\prime}(\tau)=\frac{1}{2}\left\{ 1+e^{-\tau^2}\right\}.
\end{equation}
\begin{figure}
\includegraphics[width=6cm]{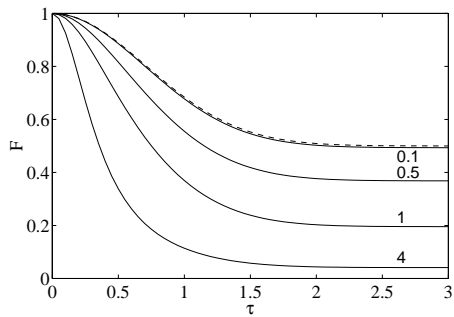}
\caption[]{Evolution of fidelities $F'(\tau)$ (dashed curve),
given by Eq. (\ref{eq:58}), and $F(\tau)$ (solid curves), given by
Eq. (\ref{eq:599}), for different values of intensity
$|\alpha|^2$. }\label{fig3}
\end{figure}
In Fig.(\ref{fig3}), we compare the fidelities for the
Eqs.(\ref{eq:599}) (\ref{eq:58}).  We find that their fidelities
are almost the same when the odd and even coherent qubit states
are in the low intensity limit. But when the intensity increases,
the fidelity given in Eq.(\ref{eq:599}) is less than that in
Eq.(\ref{eq:58}). It shows that in case of the phase damping, the
odd and even coherent states are not good coding states.

\section{conclusions and discussions }
Using a simple operator algebra solution, we have derived the
Kraus representation  of a harmonic oscillator suffering from the
zero temperature environmental effect, which is  modeled as the
amplitude damping or the phase damping.  It is worth noting that
we could not derive a compact form for the Kraus representation
when the environment is initially in the thermal field, because it
becomes very difficult to distinguish the number of photons
detected by the environment and that absorbed by the system due to
the existence of the thermal field, which can excite the system to
higher states \cite{last}.

As examples and applications of a system in a zero temperature
environment, we first give a Kraus representation of a single qubit
whose computational basis states are defined as bosonic vacuum and
single particle number states. We further discuss the environment
effect on the qubits whose computational basis states are defined
as the bosonic odd and even coherent states. We find that when the
system suffers from the amplitude damping, the loss of even number
of photons leaves the qubit, whose computational basis is defined
by the even and odd coherent states, unchanged, but the loss of an
odd number of photons changes the even or odd properties of the
qubit. If the system loses a few photons and the intensity of the
coherent states is taken infinitely large, then we can roughly say
that the loss of an even number of photons keeps qubit unchanged,
but the loss of an odd number of photons causes a bit flip error.
Such an error can be corrected by some unitary operation. But if
the computational basis is defined by the vacuum and single-photon
states, a single-photon loss is an irreversible process, we cannot
find any unitary operation to correct this error resulting from a
single photon loss. So in the amplitude damping channel, the use of
the even and odd coherent states as logical qubit is more suitable
than the use of the vacuum and single-photon state as logical
states~\cite{ptc,oliver}.

When the qubit states, whose computational basis is defined by the
even and odd coherent states, enter the phase damping channels,
the even and odd properties will not be changed, however all
off-diagonal terms of the qubit density matrices  gradually
vanish. Then the originally pure states change into mixtures of
states because of the random scattering of the environment on the
system even without loss of photons. It is much more difficult to
correct such an error. So the use of  the even and odd coherent
states as logical qubit states cannot solve the problem of the
phase damping.

\section{acknowledgments}
Yu-xi Liu thanks Franco Nori for his helpful discussions.  Yu-xi
Liu was supported by the Japan Society for the Promotion of
Science (JSPS). This work is partially supported by a Grant-in-Aid
for Scientific Research (C) (15340133) by the Japan Society for
the Promotion of Science (JSPS).

\end{document}